\documentclass[english]{article}
\usepackage[T1]{fontenc}
\usepackage[latin9]{inputenc}
\usepackage[letterpaper]{geometry}
\usepackage{color}
\usepackage{amsmath}
\usepackage{graphicx}
\usepackage{amssymb}
\usepackage{esint}

\makeatletter
\newcommand{\lyxaddress}[1]{
\par {\raggedright #1
\vspace{1.4em}
\noindent\par}
}

\makeatother

\usepackage{babel}

\begin{document}

\title{Milli-interacting dark matter interpretation of the direct-search
experiments}

\author{Quentin Wallemacq%
\thanks{quentin.wallemacq@ulg.ac.be%
}}

\maketitle

\lyxaddress{\begin{center}
IFPA, Dpartement AGO, Universit de Lige, Sart Tilman, 4000 Lige,
Belgium
\par\end{center}}
\begin{abstract}
We reinterpret the results of the direct searches for dark matter
in terms of milli-interacting dark particles. The model reproduces
the positive results from DAMA/LIBRA and CoGeNT and is consistent
with the absence of signal in the XENON100, CDMS-II/Ge and LUX detectors.
Dark atoms, interacting with standard atoms through a kinetic mixing
between photons and dark photons and a mass mixing of $\sigma$ mesons
with dark scalars, diffuse elastically in terrestrial matter where
they deposit all their energy. Reaching underground detectors through
gravity at thermal energies, they form bound states with nuclei of
the active medium by radiative capture, which causes the emission
of photons that produce the observed signals. The parameter space
of the model is explored and regions reproducing the results at the
$2\sigma$ level are obtained for each experiment.
\end{abstract}

\section{Introduction}

Dark matter has been one of the most persistent enigmas in astrophysics
since an invisible kind of matter \textcolor{black}{was} suggested
in 1933 by Zwicky as an explanation to the missing mass between galaxies.
Nowadays, the presence of dark matter is known at all cosmological
scales and it is mostly believed that it is due to a unique species
of collisionless particles, whose nature remains a mystery. One way
to solve part of the problem is to observe directly these Weakly Interacting
Massive Particles (WIMPs) in underground detectors. Such direct searches
for dark matter have started in the late 1990s and have lead today
to stunning results. The DAMA/LIBRA \cite{Bernabei:2010mq,Bernabei:2013xsa}
and CoGeNT \cite{Aalseth:2011wp,Collar:2013} experiments both have
performed temporal analyses of their signals and confirmed the presence
of an annual modulation of the event rates with statistical significances
of $9.3\,\sigma$ and $2.8\,\sigma$ respectively. CRESST-II \cite{Angloher:2011uu},
and recently CDMS-II/Si \cite{Agnese:2013rvf}, support these results
with the observation of events in their detectors that cannot be due
to background. On the other hand, XENON100 \cite{Aprile:2012nq},
CDMS-II/Ge \cite{Ahmed:2010wy} and recently LUX \cite{Akerib:2013tjd}
exclude any detection.

The current problem is that these experiments seem to come into conflict
when their results are interpreted in terms of WIMPs producing nuclear
recoils by colliding on nuclei in the detectors, although a more precise
account for theoretical and experimental uncertainties could improve
the status of WIMPs in that field. The tensions between experiments
with positive results and the apparent incompatibility of the latter
with experiments with negative results has lead to consider other
dark matter models that could provide new frameworks to reinterpret
the data. Among these, mirror matter \cite{Foot:2012rk}, millicharged
atomic dark matter \cite{Cline:2012is}, succeeding works as Ref.~\cite{Feldman:2007wj},
O-helium dark atoms \cite{Khlopov:2010ik,Khlopov:2011me,Cudell:2012fw}
and exothermic double-disk dark matter \cite{McCullough:2013jma}
propose interesting and varied mechanisms that can reconcile part
of the experiments, but always keep contradictions with the others.
Light-mediator exchange \cite{Fornengo:2011sz} provides a viable
mechanism which is able to explain the modulation effects, but its
compatibility with the experiments with negative results is still
uncertain. One common feature of all these scenarios is the high complexity
of their dark sectors with respect to WIMPs, often reaching a phenomenology
as rich as that of our ordinary sector.

The model presented here follows this trend and keeps some aspects
of the above ones, but presents new ingredients which aim at reconciling
the experiments with positive results without contradicting those
with negative results. The dark sector is composed of two new fermions
both coupled to massless dark photons with opposite couplings and
neutral dark scalars to which is coupled one of the two species via
a Yukawa coupling. The oppositely charged dark fermions bind to form
dark hydrogenoid atoms with standard atomic sizes. Such a dark matter
candidate presents self-interactions on which constraints have been
established from the Bullet cluster and from halo shapes \cite{2002ApJ...564...60M}.
To avoid them, we follow Ref. \cite{Fan:2013yva} in which the self-interacting
part of the dark sector is reduced to at most $5\%$ of the total
dark matter mass content of the galaxy, the rest being realized by
conventional collisionless particles presenting too weak interactions
with standard particles to produce any recoil in underground detectors.
The same kind of kinetic photon-dark photon mixing as in \cite{Foot:2012rk}
and \cite{Cline:2012is} produces small effective couplings of the
dark fermions to the standard photon, the former behaving therefore
like electric millicharges interacting with electrically charged standard
particles. An additional mass mixing between the standard scalar $\sigma$
meson and the dark scalar creates an attractive interaction between
one of the two dark fermions and the standard nucleons that are coupled
to $\sigma$ in the framework of an effective Yukawa theory. The dark
atoms interact sufficiently with terrestrial matter to lose all their
energy between the surface and underground detectors, reaching them
with thermal energies. There, dark and standard nuclei form bound
states by radiative capture, causing the emission of photons that
are the sources of the observed signals.

In Ref. \cite{Wallemacq:2013hsa}, the model was introduced and a
specific set of parameters that reproduced well the results from DAMA/LIBRA
and CoGeNT and presenting no contradictions with XENON100 and CDMS-II/Ge
was given. Here, the parameter space of the model is explored to determine
the regions that reproduce DAMA/LIBRA and CoGeNT at the $2\sigma$
level and to put upper and lower limits on the different parameters,
always without contradicting the null results from XENON100 and CDMS-II/Ge,
as well as the new constraint from LUX.

\section{Dark sector}

In this model, the complex part of the dark sector is realized by
two kinds of fermions, $F$ and $G$, of masses $m_{F}$ and $m_{G}$,
interacting through a dark $U(1)$ gauge interaction carried out by
dark massless photons $\Gamma$. In addition, the species $F$ exchanges
dark neutral scalars $S$ of mass $m_{S}$ via a Yukawa coupling,
which leads to the dark interaction Lagrangian: \begin{equation}
\mathcal{L}_{int}^{dark}=e'\overline{\psi}_{F}\gamma^{\mu}A'_{\mu}\psi_{F}-e'\overline{\psi}_{G}\gamma^{\mu}A'_{\mu}\psi_{G}+g'\phi_{S}\overline{\psi}_{F}\psi_{F},\label{eq:1}\end{equation}
where $\psi_{F}$ and $\psi_{G}$ are the fermionic fields of $F$
and $G$, $A'$ and $\phi_{S}$ are the vectorial and real scalars
fields of $\Gamma$ and $S$, $+e'$ and $-e'$ are the electric charges
of $F$ and $G$, and $g'$ is the Yukawa coupling of $F$ to $S$.

In order to produce nongravitational interactions between the standard
and dark sectors, we postulate that the dark photons $\Gamma$ are
kinetically mixed with the standard photons $\gamma$ and that the
dark scalars $S$ are mixed with the neutral scalar mesons $\sigma$
via a mass term, with the mixing Lagrangian: \begin{equation}
\mathcal{L}_{mix}=\frac{1}{2}\tilde{\epsilon}\mathcal{F}^{\mu\nu}\mathcal{F}'_{\mu\nu}+\tilde{\eta}\left(m_{\sigma}^{2}+m_{S}^{2}\right)\phi_{\sigma}\phi_{S},\label{eq:2}\end{equation}
 where $\mathcal{F}$ and $\mathcal{F}'$ are the electromagnetic-field-strength
tensors of the massless standard and dark photons, $\phi_{\sigma}$
is the real scalar field of the $\sigma$ meson and $m_{\sigma}=600$
MeV \cite{Amsler:2008zzb} is its mass. $\tilde{\epsilon}$ and $\tilde{\eta}$
are the dimensionless parameters of the kinetic $\gamma-\Gamma$ and
mass $\sigma-S$ mixings and are assumed to be small compared with
unity.

In principle, the model contains seven free parameters, $m_{F}$,
$m_{G}$, $m_{S}$, $e'$, $g'$, $\tilde{\epsilon}$ and $\tilde{\eta}$,
but these can in fact be reduced to four. Indeed, only the products
$\tilde{\epsilon}e'$ and $\tilde{\eta}g'$ will be directly constrained
by the direct-search experiments, which suggests to define them in
terms of the charge of the proton $e$ and of the Yukawa coupling
constant of the nucleon to the $\sigma$ meson $g=14.4$ \cite{Erkol:2005jz}:
$\tilde{\epsilon}e'\equiv\epsilon e$ and $\tilde{\eta}g'\equiv\eta g$,
where $\epsilon$ and $\eta$ are dimensionless mixing parameters.
Morever, the oppositely charged fermions $F$ and $G$ will bind to
form hydrogenoid dark atoms where $F$ and $G$ will respectively
play the roles of dark nucleus and dark electron, satisfying $m_{G}\ll m_{F}$.
The Bohr radius of such atoms is given by $a'_{0}=\frac{1}{m_{G}\alpha'}$,
where $\alpha'=\frac{e'^{2}}{4\pi}$, and gives another parameter
that will be fixed to $1$ \AA~so that dark atoms have the same size
as standard ones and can thermalize in the Earth before reaching the
underground detectors, as will be specified in\textcolor{red}{{} }\textcolor{black}{Section
\ref{sub:Interactions-of-dark}} . As a result, the parameters of
the model can be reexpressed as $m_{F}$, $m_{S}$, $\epsilon$ and
$\eta$.

The dark particles $F$ will bind to nuclei in \textcolor{black}{underground}
detectors and have therefore to be sufficiently massive to form bound
states. For that reason, we will explore masses of $F$ between $10$
GeV and $10$ TeV. The mass mixing term in \eqref{eq:2} induces an
attractive interaction between $F$ and nucleons with a range determined
by $m_{S}^{-1}$. \textcolor{black}{It cannot be too long ranged but
it must allow the existence of nucleus-$F$ bound states of at least
the size of the nucleus, so we will seek masses of $S$ between $100$
keV and $10$ MeV.}\textcolor{red}{{} }\textcolor{black}{The model parameters
that we will consider are therefore: }\[
    \begin{array}{c}
       10 $ GeV $ \leq\ m_F \leq\ 10 $ TeV$ \\
       100 $ keV $ \leq\ m_S \leq\ 10 $ MeV$ \\
       \epsilon,\eta \ll\ 1 \\
       a_0' = 1 $ \AA $
    \end{array}
\]

Note that the galactic dark matter halo could also be populated by
dark ions $F$ and $G$, but Ref. \cite{McDermott:2010pa} ensures
that if $\epsilon>9\times10^{-12}$ ($m_{F,G}/$GeV), they have been
evacuated from the disk by supernovae shock waves while galactic magnetic
fields prevent them from reentering. This condition will clearly be
satisfied by the parameters used to reproduce the results of the direct-dark-matter-search
experiments and we can consider their signals to be fully due to dark
atoms.

\section{Interaction potentials with standard matter}

Because of the mixings present in \eqref{eq:2}, the dark fermions
$F$ and $G$ can interact with our standard particles. The kinetic
$\gamma-\Gamma$ mixing induces small effective couplings $\pm\tilde{\epsilon}e'=\pm\epsilon e$
to the standard photon for $F$ and $G$. The dark species can therefore
interact electromagnetically with any charged standard particle with
millicharges $\pm\epsilon e$. Similarly, the $\sigma-S$ mass mixing
generates an effective coupling between $F$ and $\sigma$, making
$F$ capable of interacting with any standard particle coupled to
$\sigma$, i.e. the nucleons in the framework of an effective Yukawa
theory. In the nonrelativistic limit, these couplings give rise to
interaction potentials between the dark and the standard particles.

\subsection{Interactions of $F$ and $G$ fermions with nucleons and electrons}

At the elementary level, the kinetic $\gamma-\Gamma$ mixing produces
a Coulomb interaction potential between the millicharged dark particles
and the proton and the electron: \begin{equation}
V_{k}=\pm\frac{\epsilon\alpha}{r},\label{eq:3}\end{equation}
 where $k$ refers to kinetic and $\alpha=\frac{e^{2}}{4\pi}$ is
the fine structure constant. The plus and minus signs stand respectively
for the paires proton-$F$ or electron-$G$ and proton-$G$ or electron-$F$.

Since the mass mixing parameter $\tilde{\eta}$ is small, the attractive
interaction between $F$ and the nucleons is dominated by one $\sigma+S$
exchange, which gives: \begin{equation}
V_{m}=-\frac{\eta\left(m_{\sigma}^{2}+m_{S}^{2}\right)\beta}{r}\left(\frac{e^{-m_{\sigma}r}-e^{-m_{S}r}}{m_{S}^{2}-m_{\sigma}^{2}}\right),\label{eq:4}\end{equation}
 where $m$ stands for mass and $\beta=\frac{g^{2}}{4\pi}=16.5$.
Note that because $m_{S}\ll m_{\sigma}$, the potential \eqref{eq:4}
is essentially a Yukawa potential of range $m_{S}^{-1}$: $V_{m}\simeq-\frac{\eta\beta}{r}e^{-m_{S}r}$.

\subsection{Interactions of $F$ fermions with nuclei}

Because of their interactions with nucleons, the dark fermions $F$
interact with atomic nuclei. Assuming that a nucleus of mass number
$A$ and atomic number $Z$ is a uniformly charged sphere of radius
$R=r_{0}A^{1/3}$ and volume $V=\frac{4}{3}\pi R^{3}$, the integrations
of the elementary potentials \eqref{eq:3} and \eqref{eq:4} over
its electric and nuclear charge distributions $\rho_{k}=\frac{Ze}{V}$
and $\rho_{m}=\frac{Ag}{V}$ give: \begin{equation}
\begin{array}{ccl}
V_{k}^{nucl}\left(r\right) & = & \int_{V}\left(V_{k}\left(\left|\vec{r}-\vec{r'}\right|\right)/e\right)\rho_{k}d\vec{r'}\\
 & = & \frac{\epsilon Z\alpha}{2R}\left(3-\frac{r^{2}}{R^{2}}\right),\,\,\,\, r<R\\
 & = & \frac{\epsilon Z\alpha}{r},\,\,\,\,\,\,\,\,\,\,\,\,\,\,\,\,\,\,\,\,\,\,\,\,\,\,\,\,\,\,\, r>R\end{array}\label{eq:5}\end{equation}
 and $V_{m}^{nucl}\left(r\right)=\int_{V}\left(V\left(\left|\vec{r}-\vec{r'}\right|\right)_{m}/g\right)\rho_{m}d\vec{r'}$,
i.e. \begin{equation}
\begin{array}{ccl}
V_{m}^{nucl}\left(r<R\right) & = & -\frac{V_{0}}{r}\left[2r\left(m_{\sigma}^{-2}-m_{S}^{-2}\right)+\left(R+m_{\sigma}^{-1}\right)m_{\pi}^{-2}\left(e^{-m_{\sigma}r}-e^{m_{\sigma}r}\right)e^{-m_{\sigma}R}\right.\\
\\ &  & \left.-\left(R+m_{S}^{-1}\right)m_{S}^{-2}\left(e^{-m_{S}r}-e^{m_{S}r}\right)e^{-m_{S}R}\right],\\
\\V_{m}^{nucl}\left(r>R\right) & = & -\frac{V_{0}}{r}\left[m_{\sigma}^{-2}e^{-m_{\sigma}r}\left(e^{m_{\sigma}R}\left(R-m_{\sigma}^{-1}\right)+e^{-m_{\sigma}R}\left(R+m_{\sigma}^{-1}\right)\right)\right.\\
\\ &  & \left.-m_{S}^{-2}e^{-m_{S}r}\left(e^{m_{S}R}\left(R-m_{S}^{-1}\right)+e^{-m_{S}R}\left(R+m_{S}^{-1}\right)\right)\right],\end{array}\label{eq:6}\end{equation}
 where $nucl$ indicates nucleus, $\vec{r'}$ is the position vector
of a charge element in the nucleus, $V_{0}=3\eta\left(m_{\sigma}^{2}+m_{S}^{2}\right)\beta/\left(2r_{0}^{3}\left(m_{S}^{2}-m_{\sigma}^{2}\right)\right)$
and $r_{0}=1.2$ fm.

$V_{k}^{nucl}$ consists in a repulsive Coulomb potential outside
the nucleus and in a harmonic potential, which is concave down, inside.
Both are continuously connected at $r=R$ (as well as their first
derivatives), with an inflection point at $r=R$ and a maximum reached
at $r=0$, where the first derivative is zero.

$V_{m}^{nucl}$ corresponds to a finite attractive well, with a size
of the order of $m_{S}^{-1}$, an inflection point at $r=R$ and tending
to zero as $\frac{-1}{r}e^{-m_{S}r}$ outside the nucleus.

The total nucleus-$F$ potential $V_{k}^{nucl}+V_{m}^{nucl}$ is therefore
a negative attractive well at distances $r\lesssim m_{S}^{-1}$ that
is continuously connected, together with its first derivative, to
a positive potential barrier, coming from the repulsive Coulomb potential,
at larger distances. As $r\rightarrow\infty$, $V_{m}^{nucl}$ rapidly
tends to zero and the total potential is dominated by the Coulomb
part. In order to reproduce the direct-search experiments, the depth
of the well, mainly determined by the parameter $\eta$, will be of
the order of $10$ keV while the barrier, which height depends on
$m_{S}$ and $\epsilon$, will rise up to a few eV.

\subsection{Interactions of dark $FG$ atoms with terrestrial atoms\label{sub:Interactions-of-dark}}

The galactic dark atoms interact, after hitting the surface of the
Earth because of its motion in the dark matter halo, with terrestrial
atoms under the surface. Because $m_{F}\gg m_{G}$, $F$ plays the
role of a dark nucleus while $G$ is spherically distributed around
it, so that the mass of the bound state $m_{FG}$ is almost equal
to $m_{F}$. To model the atom-dark atom interaction, the dark atoms,
as well as the terrestrial ones, are seen as uniformly charged spheres
of charges $-\epsilon e$ and $-Ze$ and radii $a'_{0}$ and $a_{0}$
respectively, where $Z$ is the atomic number of the terrestrial atom,
with opposite pointlike charges at their centers. The atomic radii
are both fixed to $1$ \AA~ to allow sufficient interaction for the
dark atoms to lose all their kinetic energy by elastic collisions
between the surface and the underground detectors and hence to reach
them with thermal energies.

The atom-dark atom interaction potential is then the sum of the electrostatic
interaction $V_{k}^{at}$ and the $\sigma+S$ echange $V_{m}^{at}$
between the dark nucleus $F$ and the atomic nucleus. $V_{k}^{at}$
is obtained by adding the contributions from the four pairs of crossed
substructures: nucleus-$F$ (pointlike-pointlike, pure Coulomb repulsion),
nucleus-$G$ distribution (pointlike-sphere, attractive of form \eqref{eq:5}),
electron distribution-$F$ (sphere-pointlike, attractive of form \eqref{eq:5})
and electron distribution-$G$ distribution (sphere-sphere, obtained
by integrating the form \eqref{eq:5} over a uniformly charged sphere
which center is separated by a distance $r$ from the center of the
first one). This gives: \begin{equation}
\begin{array}{ccl}
V_{k}^{at} & = & \frac{\epsilon Z\alpha}{160a_{0}^{6}}\left(-r^{5}+30a_{0}^{2}r^{3}+80a_{0}^{3}r^{2}-288a_{0}^{5}+\frac{160a_{0}^{6}}{r}\right),\,\,\,\,\,\,\,\,\,\,\,\,\,\,\,\, r<a_{0}\\
 & = & \frac{\epsilon Z\alpha}{160a_{0}^{6}}\left(-r^{5}+30a_{0}^{2}r^{3}-80a_{0}^{3}r^{2}+192a_{0}^{5}-\frac{160a_{0}^{6}}{r}\right),\,\, a_{0}<r<2a_{0}\\
 & = & 0,\,\,\,\,\,\,\,\,\,\,\,\,\,\,\,\,\,\,\,\,\,\,\,\,\,\,\,\,\,\,\,\,\,\,\,\,\,\,\,\,\,\,\,\,\,\,\,\,\,\,\,\,\,\,\,\,\,\,\,\,\,\,\,\,\,\,\,\,\,\,\,\,\,\,\,\,\,\,\,\,\,\,\,\,\,\,\,\,\,\,\,\,\,\,\,\,\,\,\,\,\,\,\,\,\,\,\,\,\,\,\,\,\,\,\,\,\,\,\,\,\,\,\,\,\,\,\,\,\,\,\,\,\,\,\,\,\,\,\, r>2a_{0}\end{array}\label{eq:7}\end{equation}
where the upper label $at$ refers to atomic. Because the nucleus
is here supposed to be pointlike, $V_{m}^{at}$ is simply obtained
by multiplying \eqref{eq:4} by the number $A$ of nucleons in the
nucleus: \begin{equation}
V_{m}^{at}\left(r\right)=-\frac{\eta\left(m_{\sigma}^{2}+m_{S}^{2}\right)A\beta}{r}\left(\frac{e^{-m_{\sigma}r}-e^{-m_{S}r}}{m_{S}^{2}-m_{\sigma}^{2}}\right),\label{eq:8}\end{equation}

The atom-dark atom electrostatic potential $V_{k}^{at}$ shows three
parts as a function of the distance $r$ between the two centers.
Each sphere appears neutral from outside because the positive charge
at the center compensates exactly the negative charge distributed
in the sphere, so that there is no interaction when they are completely
separated ($r>2a_{0}$). As they merge ($a_{0}<r<2a_{0})$, the electrostatic
potential becomes attractive due to the attraction between the nucleus
of each sphere and the negatively charged distribution of the other
one. When the nuclei enter simultaneously in the approaching spheres
($r=a_{0}$), an inflection point accurs and the potential reaches
then a minimum ($r=0.88$ \AA~ for $a_{0}=1$ \AA~). As the centers
continue to approach each other ($r<0.88$ \AA~ for $a_{0}=1$ \AA~),
the potential becomes repulsive due to the Coulomb repulsion between
nuclei. The potential well appearing when the two atoms merge will
have a depth of the order of $10^{-4}-10^{-3}$ eV and will therefore
not contain any bound state (or if any, not thermally stable), so
that it will not contribute in the following to the formation of atom-dark
atom bound states.

As $m_{S}\ll m_{\sigma},$ $V_{m}^{at}\simeq-\frac{\eta A\beta}{r}e^{-m_{S}r}$,
which is a pure Yukawa potential. The total atom-dark atom potential
$V_{k}^{at}+V_{m}^{at}$ is therefore essentially equal to its electrostatic
part $V_{k}^{at}$ for $r<m{}_{S}^{-1}$, while the attractive part
$V_{m}^{at}$ dominates at smaller distances.

\section{From space to underground detectors}

\subsection{Thermalization in the terrestrial crust}

Due to its orbital motion around the Sun, which turns around the center
of the galaxy, the Earth moves through the galactic dark matter halo.
This results in a wind of dark atoms hitting the surface of the Earth
throughout the year.%
\footnote{However, according to Ref. \cite{Fan:2013yva}, it is expected that
the subdominant self-interacting species form a disk rotating around
the galactic center, so that the incident flux on Earth might be different
than in the halo assumption.%
} A dark atom penetrates the surface and starts interacting with terrestrial
atoms via the atomic potentials \eqref{eq:7} and \eqref{eq:8}. As
there is no stable bound state in the total atomic potential with
the relatively light terrestrial atoms, the diffusions are purely
elastic. If the elastic diffusion cross section is sufficiently large,
then the dark atom can deposit all its energy in the terrestrial matter,
assumed to be mainly made of silicon atoms with $Z_{Si}=14$ and $A_{si}=28$,
before reaching an undergound detector typically located at a depth
of $1$ km.

A differential elastic diffusion cross section $\frac{d\sigma}{d\Omega}$
deriving from a two-body-interaction potential $V\left(\vec{r}\right)$
can be obtained in the framework of the Born approximation via the
Fourier transform of the potential: $\frac{d\sigma}{d\Omega}=\frac{\mu^{2}}{4\pi^{2}}\left|\int d\vec{r}e^{-i\vec{K}.\vec{r}}V\left(\vec{r}\right)\right|^{2}$,
where $\mu$ is the reduced mass of the two-body system and $\vec{K}$
is the transferred momentum. Here, from potentials \eqref{eq:7} and
\eqref{eq:8} and in the center-of-mass frame of the silicon-$F$
system, we get: \begin{equation}
\begin{array}{ccc}
\left(\frac{d\sigma}{d\Omega}\right)^{at} & = & \left(\frac{d\sigma}{d\Omega}\right)_{k}^{at}+\left(\frac{d\sigma}{d\Omega}\right)_{m}^{at}-\frac{4\mu^{2}\epsilon\eta Z_{Si}A_{Si}\alpha\beta}{a_{0}^{6}}\left(\frac{m_{\sigma}^{2}+m_{S}^{2}}{m_{S}^{2}-m_{\sigma}^{2}}\right)\frac{I}{K^{8}}\left[\frac{1}{m_{\sigma}^{2}+K^{2}}-\frac{1}{m_{S}^{2}+K^{2}}\right]\end{array},\label{eq:9}\end{equation}
 with \begin{equation}
\begin{array}{ccl}
\left(\frac{d\sigma}{d\Omega}\right)_{k}^{at} & = & \frac{\mu^{2}\epsilon^{2}Z_{mSi}^{2}\alpha^{2}}{a_{0}^{12}}\frac{1}{K^{16}}I^{2},\\
\\I & = & 9\left(K^{2}a_{0}^{2}+1\right)+9\cos\left(2Ka_{0}\right)\left(K^{2}a_{0}^{2}-1\right)+12\cos\left(Ka_{0}\right)K^{4}a_{0}^{4}\\
\\ &  & -18\sin\left(2Ka_{0}\right)Ka_{0}-12\sin\left(Ka_{0}\right)K^{3}a_{0}^{3}+2K^{6}a_{0}^{6}\end{array}\label{eq:10}\end{equation}
 for the electrostatic interaction and \begin{equation}
\left(\frac{d\sigma}{d\Omega}\right)_{m}^{at}=4\mu^{2}\eta^{2}A_{m}^{2}\beta^{2}\left(\frac{m_{\sigma}^{2}+m_{S}^{2}}{m_{S}^{2}-m_{\sigma}^{2}}\right)^{2}\left[\frac{1}{m_{\sigma}^{2}+K^{2}}-\frac{1}{m_{S}^{2}+K^{2}}\right]^{2}\label{eq:11}\end{equation}
 for the $\sigma+S$ exchange. $\mu=\frac{m_{F}m_{Si}}{m_{F}+m_{Si}}$,
where $m_{Si}$ is the mass of a silicon atom, $K=2k\sin\theta/2$,
where $k=\sqrt{2\mu E}$ is the initial momentum and $\theta$ is
the deflection angle with respect to the collision axis.

For a dark atom to thermalize between the surface and an underground
detector, we have to ensure that its penetration length does not exceed
$1$ km. It is estimated by assuming a linear path of the dark atom
through terrestrial matter: \begin{equation}
x=\int_{E_{th}}^{E_{0}}\frac{dE}{\left|dE/dx\right|}<1,\label{eq:12}\end{equation}
 where $\frac{dE}{dx}$ is the energy loss per unit length in the
frame of the Earth: \begin{equation}
\frac{dE}{dx}=n_{Si}\int_{\Omega}\triangle K\left(\frac{d\sigma}{d\Omega}\right)^{at}d\Omega,\label{eq:13}\end{equation}
 obtained by integrating over all diffusion angles. In \eqref{eq:12},
the integration is performed from the initial kinetic energy of the
dark atom $E_{0}$ to the thermal energy of the medium $E_{th}=\frac{3}{2}T_{med}$,
where $T_{med}\simeq300$ K. In \eqref{eq:13}, $n_{Si}$ is the number
density of atoms in the terrestrial crust and $\triangle K=\frac{p^{2}\left(\cos\theta-1\right)}{m_{Si}}$
is the energy lost in the frame of the Earth for each collision with
a silicon atom at rest in the terrestrial surface, expressed in terms
of the momentum $p$ of each atom in the center-of-mass frame. It
is clear that the linear path approximation is valid only when $m_{F}\gg m_{Si}$,
but it gives in the other cases an upper limit on the penetration
length of a dark atom through the Earth, which is of interest here.

\subsection{Drift down towards underground detectors}

Once it has thermalized, a dark atom starts to drift towards the center
of the Earth by gravity until it reaches an underground detector.
The number density of dark atoms in the detector $n_{F}$ is determined
by the equilibrium between the infalling flux at the surface and the
down-drifting thermalized flux: $\frac{n_{0}}{4}\left|\vec{V}_{h}+\vec{V}_{E}\right|=n_{F}V_{d}$,
where $n_0$ (cm$^{-3}$)$=3\times10^{-4}/m_F$(TeV) is the local number
density of dark atoms, $\vec{V}_{h}+\vec{V}_{E}$ is the superposition
of the orbital velocity of the Sun around the galactic center $\vec{V}_{h}$
and of the Earth around the sun $\vec{V}_{E}$, and $V_{d}$ is the
drift velocity of the dark atoms in the terrestrial matter once they
have thermalized. Because of the orbital motion of the Earth around
the Sun, $\left|\vec{V}_{h}+\vec{V}_{E}\right|$ is modulated in time:
$\left|\vec{V}_{h}+\vec{V}_{E}\right|=V_{h}+V_{E}\cos\gamma\cos\left(\omega\left(t-t_{0}\right)\right)$,
so that $n_{F}$ can be written as: \begin{equation}
n_{F}=n_{F}^{0}+n_{F}^{m}\cos\left(\omega\left(t-t_{0}\right)\right),\label{eq:14}\end{equation}
 where $\gamma\simeq60^{\circ}$ is the inclination angle of the Earth
orbital plane with respect to the galactic plane, $\omega=\frac{2\pi}{T_{orb}}$
is the angular frequency of the orbital motion of the Earth, $T_{orb}=1$
yr is the orbital period and $t_{0}\simeq$ June $2$ is the period
of the year when the Earth and Sun orbital velocities are aligned.
As $V_{d}=\frac{g}{n\left\langle \sigma_{k}^{at}v\right\rangle }$,
where $g=980$ cm/s$^{2}$ is the acceleration of gravity and $n\simeq5\times10^{22}$
cm$^{-3}$ is the number density of atoms in the terrestrial crust,
the constant and modulated parts $n_{F}^{0}$ and $n_{F}^{m}$ can
be expressed as: \begin{equation}
n_{F}^{0}=\frac{n_{0}\, n\,\left\langle \sigma_{k}^{at}v\right\rangle }{4g}V_{h}\label{eq:15}\end{equation}
 and \begin{equation}
n_{F}^{m}=\frac{n_{0}\, n\,\left\langle \sigma_{k}^{at}v\right\rangle }{4g}V_{E}\cos\gamma.\label{eq:16}\end{equation}
 In these two expressions, $V_{h}=220\times10^{5}$ cm/s, $V_{E}=29.5\times10^{5}$
cm/s and $n\,\left\langle \sigma_{k}^{at}v\right\rangle $ is the
rate of elastic collisions between a thermalized dark atom $FG$ and
terrestrial atoms, averaged over a Maxwellian velocity distribution
at temperature $T_{med}\simeq300$ K. $\sigma_{k}^{at}$ is obtained
by integrating the differential cross section \eqref{eq:10} over
all diffusion angles and $v$ is the relative velocity between a dark
atom and a terrestrial atom. Note that $\sigma_{k}^{at}$ dominates
over $\sigma_{m}^{at}$ at low energies, so there is no need to consider
the total cross section $\sigma^{at}$ here.

Arriving in the detector at room temperature, a dark atom still has
to thermalize at the operating temperature. The latter is always lower
than $300$ K, except for the DAMA detectors, which operate at room
temperature. \textcolor{black}{We will check that this second thermalization
at the edge of the detector is realized over a distance much smaller
than the typical size of the device and can therefore be considered
as instantaneous.}

\subsection{Bound-state-formation events\label{sub:Bound-state-formation-events}}

In the active medium, the dark atoms undergo collisions with the constituent
atoms. Because of the Coulomb barrier due to the electric repulsion
between nuclei (potential \eqref{eq:5}), most of these collisions
are elastic but sometimes tunneling through the barrier can occur
and bring a dark nucleus $F$ into the region of the potential well
present at smaller distance, due to the exchange of $\sigma$ and
$S$ between $F$ and the nuclei of the detector (potential \eqref{eq:6}).
There, electric dipole transitions E1 produce de-excitation of the
system to low-energy bound states by emission of photons that can
be detected, causing the observed signal. At this point, only the
interaction between nuclei $V_{k}^{nucl}+V_{m}^{nucl}$ is therefore
considered to calculate the capture cross section, since it dominates
at small distance ($r\lesssim1$ \AA) and because the long-range
part of the atom-dark atom potential $V_{k}^{at}+V_{m}^{at}$ is negligible
and does not affect the initial diffusion eigenstate.

At thermal energies, to order $v/c$, only the partial $s$-wave of
an incident plane wave on an attractive center is affected by the
potential. Due to selection rules, direct E1 transitions to final
$s$-bound states are forbidden. It can also be shown that magnetic
dipole and electric quadrupole transitions M1 and E2 to such final
levels are not present \cite{Segre:1977}, leaving only the possibility
to capture $F$ in two E1 transitions, first to a $p$-bound state
and then to an s-bound state, corresponding respectively to levels
at relative angular momenta $l=1$ and $l=0$ in the nucleus-$F$
potential $V_{k}^{nucl}+V_{m}^{nucl}$. The E1 capture cross section
of $F$ by a nucleus of charge $Ze$ and mass $m$ is then given by:
\begin{equation}
\sigma_{capt}=\frac{32\pi^{2}Z^{2}\alpha}{3\sqrt{2}}\left(\frac{m_{F}}{m_{F}+m}\right)^{2}\frac{1}{\sqrt{\mu}}\frac{\left(E-E_{f}\right)^{3}}{E^{3/2}}D^{2}\mbox{,}\label{eq:17}\end{equation}
 where $\mu=\frac{m_{F}m}{m_{F}+m}$ is the reduced mass of the nucleus-$F$
system, $E$ is the total incident energy in the center-of-mass frame,
$E_{f}$ is the binding energy of the lower bound state at $l=1$
and $D=\int_{0}^{\infty}rR_{f}\left(r\right)R\left(r\right)r^{2}dr$,
where $R$ and $R_{f}$ are the radial parts of the eigenfunctions
of energies $E$ and $E_{f}$, $r$ being the relative distance between
$F$ and the nucleus.

It is important to note here that each capture event will give rise
to the emission of two photons. For the events to be seen as single-hit
events, as stated by DAMA, one will require that the first emitted
photon with the greatest possible energy, corresponding to the E1
capture from the continuum at $E\sim10^{-2}$ eV to the lower $p$-state
$E_{f}$, has an energy below the threshold $E_{threshold}$ of the
considered experiment. In other words, we will have $\left|E_{f}-E\right|\simeq\left|E_{f}\right|<E_{threshold}$.
The second emitted photon, corresponding to the E1 transition from
a $p$-state $E^{l=1}$ to an $s$-state $E^{l=0}$, will have an
energy beyong the threshold, i.e. $\left|E^{l=0}-E^{l=1}\right|>E_{threshold}$. 

Thermal motion in a detector at temperature $T$ made of nuclei $N$
gives rise to collisions between $N$ and $F$ species and hence to
the event counting rate per unit volume: \begin{equation}
R=n_{F}n_{N}<\sigma_{capt}v>,\label{eq:18}\end{equation}
where $n_{F}$ and $n_{N}$ are the number densities of $F$ and $N$
in the detector and $<\sigma_{capt}v>$ is the thermally averaged
capture cross section times the relative velocity. Using Maxwellian
velocity distributions at temperature $T$ in the frame of the detector,
passing to center-of-mass and relative velocities $\overrightarrow{v}_{CM}$
and $\overrightarrow{v}$ and performing the integral over the center-of-mass
variables, we get: \begin{equation}
R=8\pi n_{F}n_{N}\frac{1}{\left(2\pi T\right)^{3/2}}\frac{1}{\mu^{1/2}}\int_{0}^{\infty}\sigma_{capt}\left(E\right)Ee^{-E/T}dE.\label{eq:19}\end{equation}

Given the modulated form \eqref{eq:14} of the number density of $F$,
one gets a modulated expression for the event rate: \begin{equation}
R=R^{0}+R^{m}\cos\left(\omega\left(t-t_{0}\right)\right).\label{eq:20}\end{equation}
 The constant and modulated parts $R^{0}$ and $R^{m}$, when expressed
in counts per day and per kilogram (cpd/kg), are given by: \begin{equation}
\begin{array}{ccc}
R^{0} & = & Cn_{F}^{0}\int_{0}^{\infty}\sigma_{capt}\left(E\right)Ee^{-E/T}dE\\
R^{m} & = & Cn_{F}^{m}\int_{0}^{\infty}\sigma_{capt}\left(E\right)Ee^{-E/T}dE\end{array},\label{eq:21}\end{equation}
 with%
\footnote{Note that a factor of $\pi$ was missing in $C$ in Ref. \cite{Wallemacq:2013hsa}
and has been corrected here.%
}\[
C=7.54\times10^{11}\frac{QtN_{Av}}{M_{mol}}\frac{1}{(2\pi T)^{3/2}}\frac{1}{\mu^{1/2}}\]
 where $Q=1000$ g, $t=86400$ s, $N_{Av}=6.022\times10^{23}$ and
$M_{mol}$ is the molar mass of the active medium of the detector
in g/mol.

An important feature of the model is its reinterpretation of the results
of the direct-search experiments in terms of bound-state-formation
events emitting photons that produce the observed signals. This is
in opposition to the common scenario where WIMPs colliding on nuclei
at velocity $\sim220$ km/s produce nuclear recoils: here, the thermal
energies in play in the detectors are insufficient to create such
recoils, and the emitted photons cause electron recoils. In experiments
that do not discriminate between these two kinds of recoils, as DAMA/LIBRA
and CoGeNT, the reinterpretation is straightforward. In experiments
with a discrimination power, the present dark atoms are good candidates
if the results are negative, as it is the case for XENON100, LUX and
CDMS-II/Ge. Indeed, even if the bound-state-formation events cannot
be naturally suppressed, the remaining events will be interpreted
as backgrounds and rejected. Further studies have to be performed
in the case of discriminative experiments with positive results, as
CRESST-II and CDMS-II/Si, to find if it is possible that the observed
nuclear recoils may be misinterpreted bound-state-formation events
occuring near the edge of those detectors.

\section{Exploring the parameter space}

\subsection{Reproduction of the results from DAMA and CoGeNT}

The DAMA/LIBRA and CoGeNT experiments observe integrated modulation
amplitudes $\tilde{R}_{DAMA}^{m}=\left(0.0464\pm0.0052\right)$ cpd/kg
and $\tilde{R}_{CoGeNT}^{m}=\left(1.66\pm0.38\right)$ cpd/kg in the
energy intervals $\left(2-6\right)$ keV and $\left(0.5-2.5\right)$
keV respectively. 

As a first approximation and for simplicity, the signal is supposed
to be made of one monochromatic line of energy $\triangle E=E_{g}-E_{f}$,
where $E_{g}$ is the ground state at $l=0$, falling within the detection
range.%
\footnote{It would be interesting to reproduce the observed energy spectra by
taking into account the different possible transitions from the $p$-states
to the $s$-states.%
} 

The $4$-dimensional parameter space of the model is explored separately
for DAMA and CoGeNT in order to reproduce the observed rates and energy
intervals at the $2\sigma$ level, which gives corresponding regions
for each experiment. We use the isotopes $^{127}$I and $^{74}$Ge
respectively for DAMA and CoGeNT, as their detectors are made of NaI
and Ge crystals. The choice of the iodine component of the DAMA/LIBRA
experiment, rather than $^{23}$Na, is crucial since it allows to
get rid of the formation of bound states with light elements, thus
preventing the formation of anomalous heavy isotopes on Earth and
during Big Bang Nucleosynthesis. A direct consequence is that the
collisions in the terrestrial crust are purely elastic.

For each set of parameter and for each experiment, the Schrdinger
equation independent on time, with potential $V_{k}^{nucl}+V_{m}^{nucl}$
applied to the constituent nucleus, is first solved through the WKB
approximation. This gives good approximations for the eigenvalues
and eigenfunctions of the corresponding nucleus-$F$ systems, the
former allowing us to calculate $\triangle E$. The modulated part
$n_{F}^{m}$ of the number density of $F$ in the detector is then
computed using \eqref{eq:16} before finally evaluating the modulated
part of the event rate $R^{m}$ from \eqref{eq:21}, at the operating
temperatures $T=300$ K for DAMA and $T=77$ K for CoGeNT. To compute
the capture cross section $\sigma_{capt}$, given by \eqref{eq:17},
at a given energy $E$ in the center-of-mass frame of the nucleus-$F$
system, one numerically solves the radial Schordinger equation in
the continuum to get the radial part $R\left(r\right)$ of the initial
diffusion eigenstate and calculate the matrix element $D$ of the
electric dipole operator.

\begin{figure}
\begin{centering}
\includegraphics[scale=0.55]{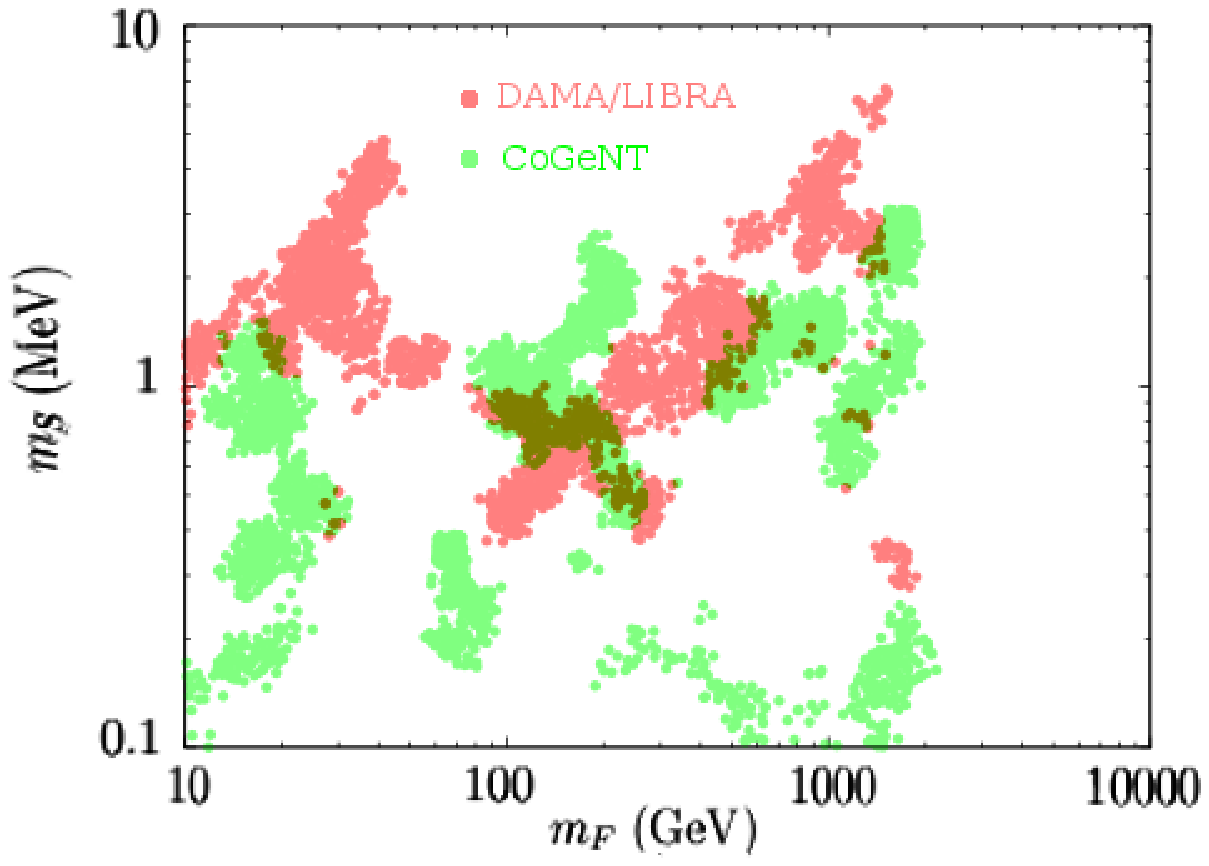}\includegraphics[scale=0.55]{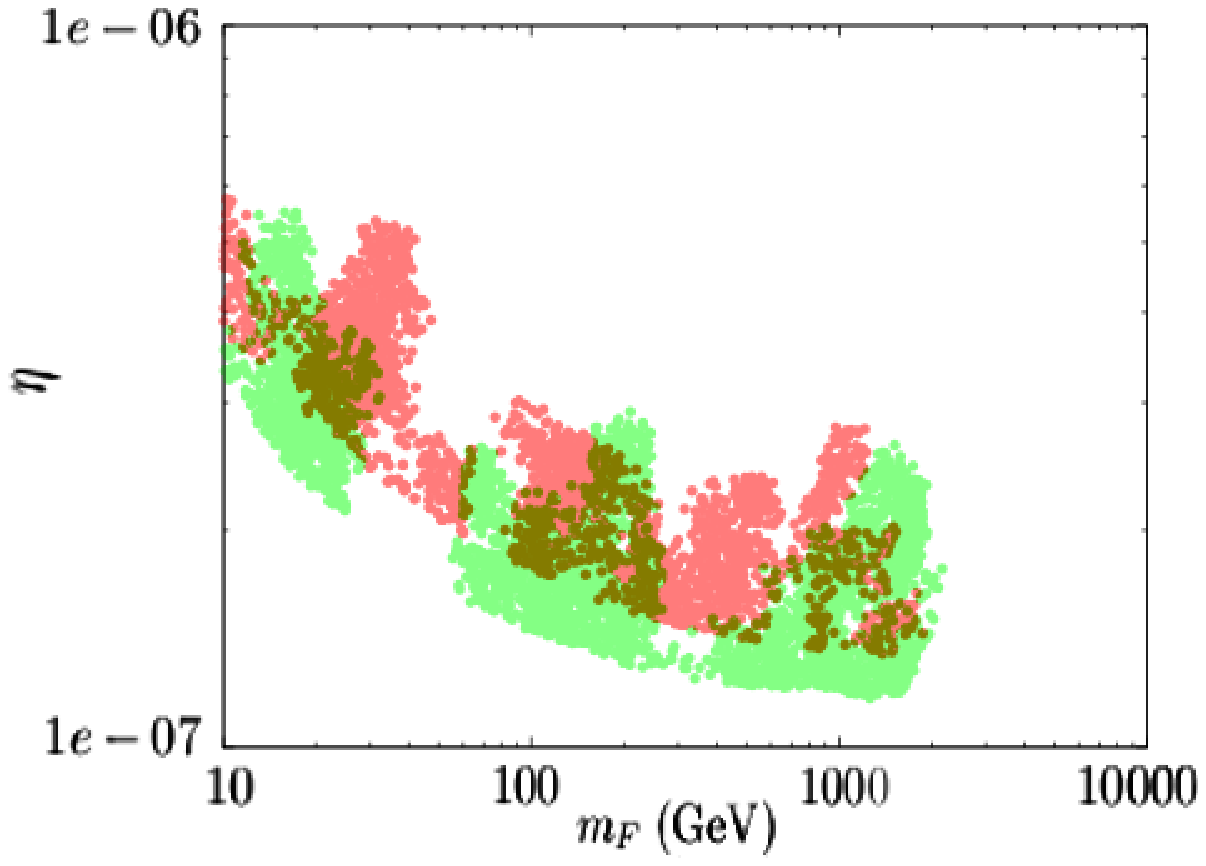}
\par\end{centering}

\begin{centering}
\includegraphics[scale=0.55]{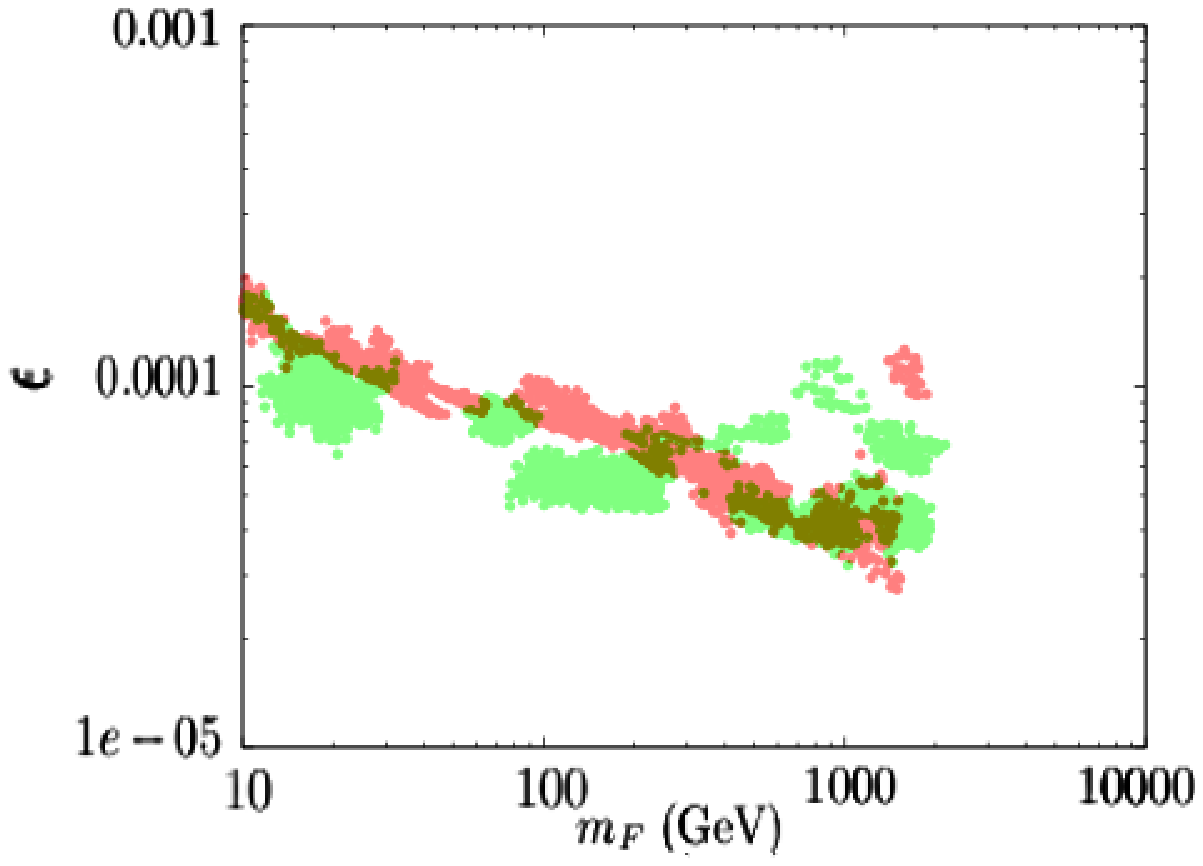}\ \ \includegraphics[scale=0.55]{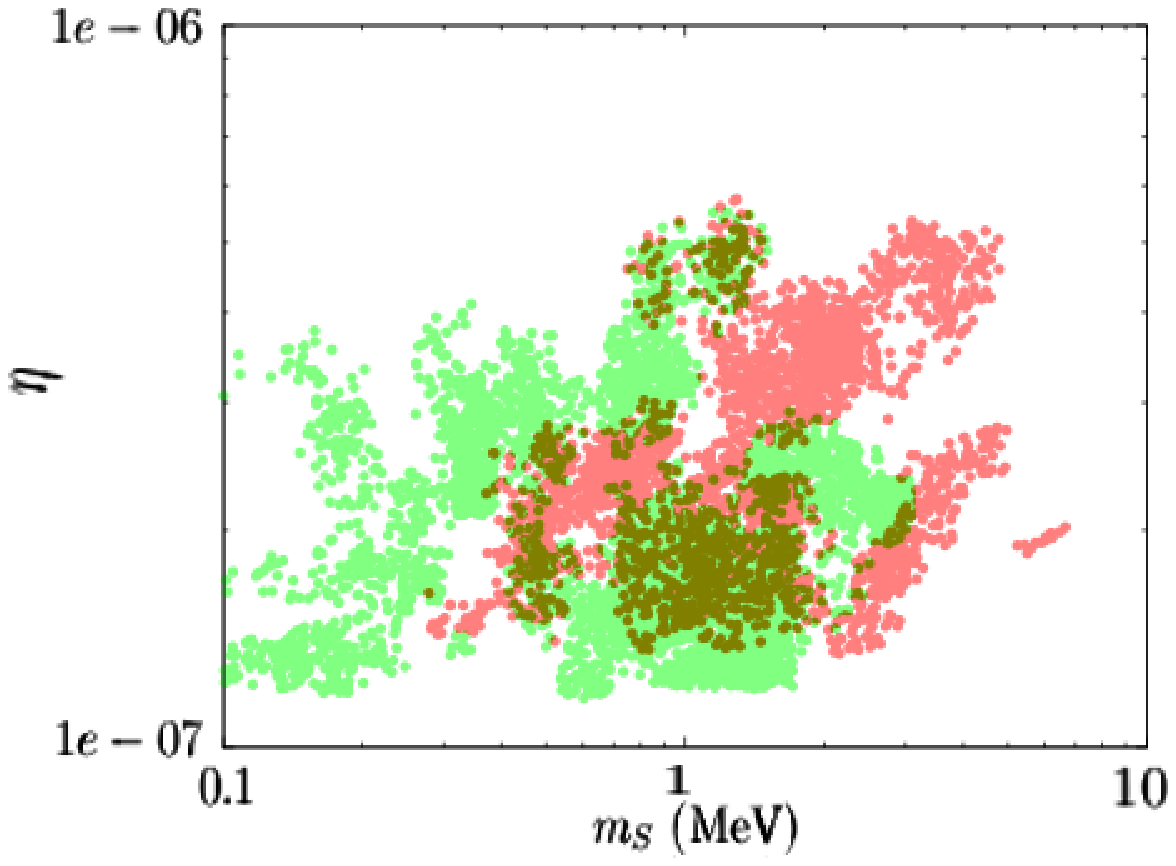}
\par\end{centering}

\begin{centering}
\includegraphics[scale=0.55]{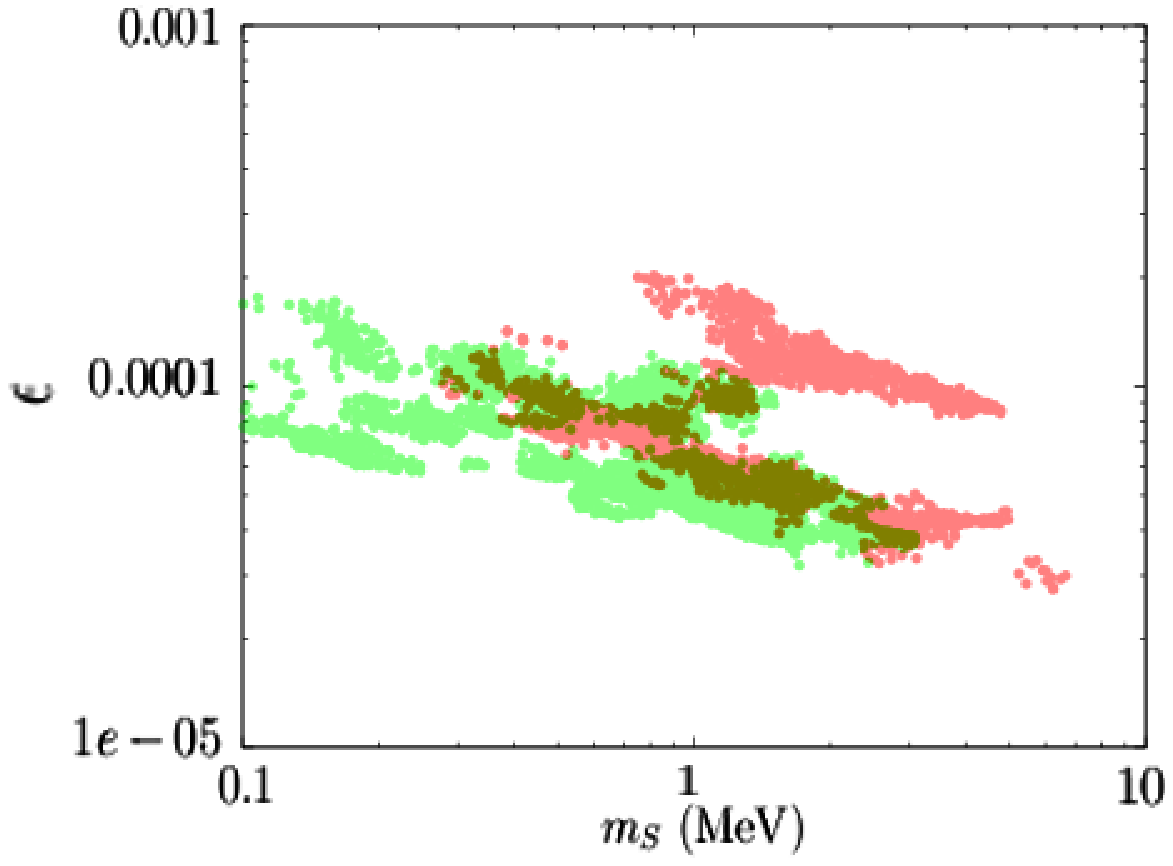}\ \ \ \ \includegraphics[scale=0.55]{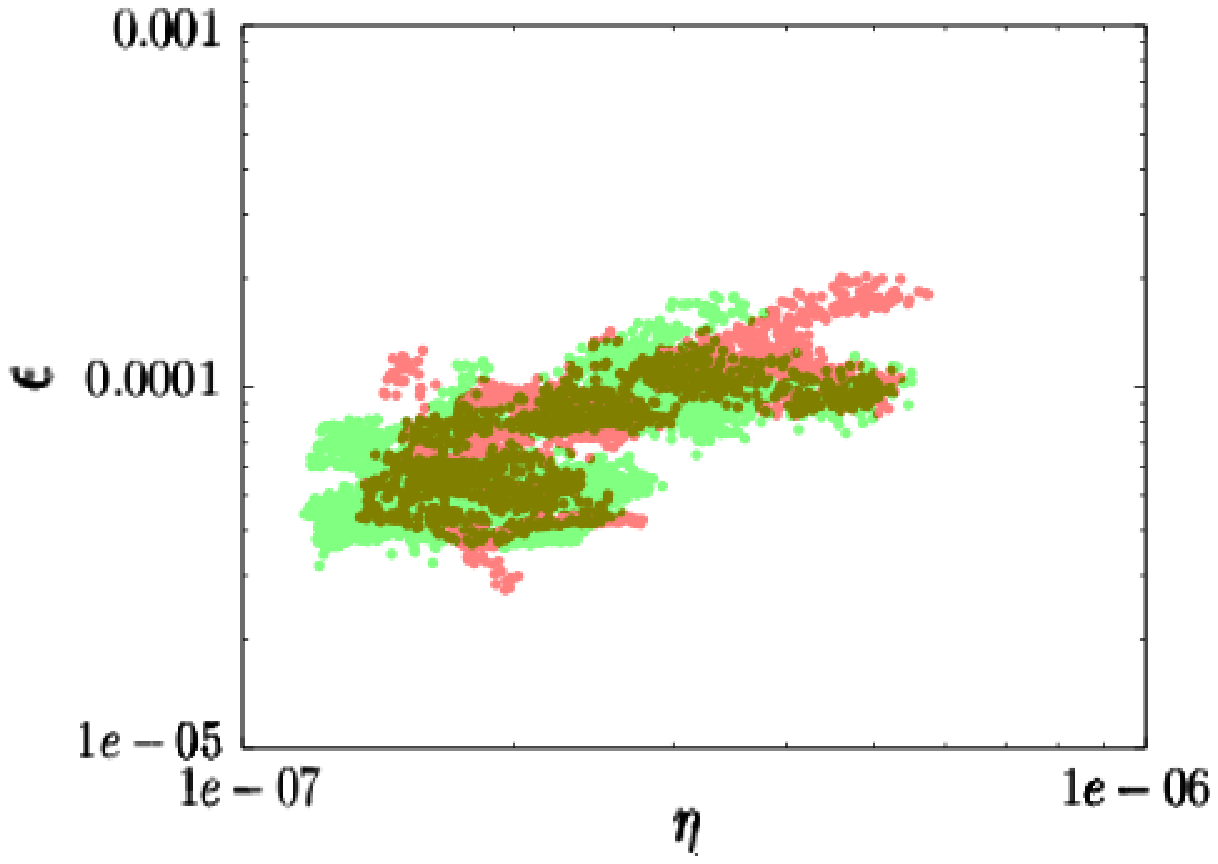}
\par\end{centering}

\caption{Two-dimensional parameter regions reproducing the DAMA/LIBRA (light
red) and CoGeNT (light green) results at the $2\sigma$ level. The
overlapping regions stand out in dark green. Top left: $\left(m_{F},m_{S}\right)$
plane. Top right: $\left(m_{F},\eta\right)$ plane. Center left: $\left(m_{F},\epsilon\right)$
plane. Center right: $\left(m_{S},\eta\right)$ plane. Bottom left:
$\left(m_{S},\epsilon\right)$ plane. Bottom right: $\mbox{\ensuremath{\left(\eta,\epsilon\right)}}$
plane.}
\label{Flo:regions}
\end{figure}

The regions are projected in two dimensions by combining all the possible
pairs of parameters and are given in Figure \ref{Flo:regions}. For
each model, one has ensured that the first emitted photon has an energy
below the threshold of the considered experiment while the second
one has an energy beyond the threshold, that thermalization occurs
before $1$ km, that no bound states can form with elements characterized
by $Z\leq14$ ($Z=14$ being silicon), and that thermalization at
the edge of the CoGeNT detector requires a penetration length much
shorter than the size of the detector. For the latter point, we have
used \eqref{eq:12} and \eqref{eq:13} with $E_{0}=\frac{3}{2}T_{room}$,
where $T_{room}=300$ K is the initial room temperature, and $E_{th}=\frac{3}{2}T$,
where $T=77$ K is the final temperature.

From the overlapping regions in the projected parameter spaces implying
$m_{F}$, we see that possible values for that parameter are between
$10$ GeV and $2$ TeV. The upper limit comes from the requirement
that the penetration length must be less than $1$ km. Analyzing the
regions where the parameter $m_{S}$ is involved indicates that the
values reproducing both the DAMA/LIBRA and CoGeNT experiments at the
$2\sigma$ level must lie within $\left[0.4,3\right]$ MeV. In the
same way, we find that $\eta$ ranges from $1.3\times10^{-7}$ to
$5\times10^{-7}$ while $\epsilon$ goes from $3\times10^{-5}$ to
$2\times10^{-4}$.

\subsection{Considerations about the constraints on $\eta$ and $\epsilon$}

One has derived, in Ref. \cite{Wallemacq:2013hsa}, a constraint on
$\tilde{\eta}=\frac{\eta g}{g'}$ from unseen vector meson disintegrations:
$\tilde{\eta}<1.2\times10^{-4}$. In principle, it is not applicable
to $\eta$, but a reasonable choice would consist in posing $g'=g$.
In this case, the constraint translates directly to $\eta$ and we
see that all the previous models satisfy it easily, by two or three
orders of magnitude.

The cosmological and astrophysical constraints on $\epsilon$, generally
derived in the framework of models with a single millicharged species
realizing the full cosmological dark matter density, cannot be applied
directly to this subdominant, atomic and millicharged scenario, and
should in any case be somewhat weakened. However, constraints from
accelerators can always be used. For masses $m_{F}\geq1$ GeV, they
let a large allowed window for $\epsilon<0.1$ \cite{Jaeckel:2010ni},
which is clearly the case here. Some interesting discussion may arise
from the lighter species $G$, constraints on $\epsilon$ from accelerators
being stronger for smaller masses. Similarly to $\eta$, $m_{G}$
is not directly constrained by the previous analysis but only the
product of $m_{G}$ and $e'^{2}$ through the Bohr radius $a'_{0}$
of the dark atoms. However, if we do once again the reasonable assumption
$e'\simeq e$, then the adopted value of $a'_{0}=1$~\AA~ leads to
$m_{G}\simeq m_{e}$, where $m_{e}$ is the mass of the electron.
It turns out that for $m_{G}\sim1$ MeV, the upper limit on $\epsilon$
from accelerators lies just in the interval deduced in the previous
section from direct experiments. If it is so, we could therefore be
close to a discovery of millicharges in accelerators via the component
$G$.

\subsection{Consistency with XENON100, CDMS-II/Ge and LUX}

For the models of Figure \ref{Flo:regions} to be fully acceptable,
we have to ensure that they satisfy the constraints set by the experiments
that do not observe any signal, as XENON100, CDMS-II/Ge and LUX. These
are able to discriminate between nuclear and electron recoils and,
as already mentioned at the end of Section \ref{sub:Bound-state-formation-events},
bound-state-formation events producing electron recoils in such detectors
will be considered as backgrounds. Therefore, if some events remain,
they should still have a smaller rate than the observed bakground.

XENON100 and LUX have similar detectors, but LUX puts the strongest
\textcolor{black}{constraint} with expected and observed electron-recoil
backgrounds respectively of $\left(2.6\pm0.2_{stat}\pm0.4_{syst}\right)\times10^{-3}$
and $\left(3.1\pm0.2_{stat}\right)\times10^{-3}$ cpd/kg/keV$_{ee}$
in the $\left(0.9-5.3\right)$ keV$_{ee}$ range. This leaves the
possibility of an additional contribution to the expected background
of at most $5.72\times10^{-3}$ cpd/kg in that energy interval. Computing
the constant part $R^{0}$ of the rate from \eqref{eq:21} for $^{132}Xe$
and at the operating temperature $T=173$ K, and rejecting the models
leading to higher rates, does not change the ranges of parameters
previously found from the reproduction of the experiments with positive
results.

Finally, this model predicts strongly suppressed event rates in cryogenic
detectors, such as CDMS-II, where temperatures $\sim1$ mK give rise
to much too low thermal energies for the dark atoms to tunnel through
the Coulomb barrier and be captured. The rates computed~with $^{74}Ge$
at $T=1$ mK are effectively consistent with zero and are therefore
in agreement with the negative results from CDMS-II/Ge.

\section{Conclusions}

We have explored the parameter space of our milli-interacting dark
matter model and found, while only one model was given in Ref. \cite{Wallemacq:2013hsa},
that regions reproducing the direct-dark-matter-search experiments
at the $2\sigma$ level can be identified. The overlaps of the regions
of DAMA/LIBRA and CoGeNT indicate that the interesting models must
lie in the ranges $10$ GeV $\leq m_F \leq 2$ TeV, $0.4$ MeV $\leq m_S \leq 3$ MeV,
$10^{-7} \leq \eta \leq 5 \times 10^{-7}$ and $3 \times 10^{-5} \leq \epsilon \leq 2 \times 10^{-4}$.
Within these intervals, models that do not contradict the negative
results from XENON100 and LUX exist, and their rates contribute to
the expected electron-recoil background. The model naturally prevents
any bound-state-formation event in cryogenic detectors ($T\sim1$
mK), which is in agreement with the Germanium detector of CDMS-II.
Some difficulties appear however with the CRESST-II and CDMS-II/Si
cryogenic experiments, for which the collisions at the edges of the
detectors should be studied in detail, when the particles are still
at room temperature and can have sufficient energies to be captured
and produce a signal.

More than giving constraints on the parameters of a specific model,
it has been shown here that it is possible, in the framework of dark
matter models containing a sector with a richness and a complexity
similar to ours, to reconcile experiments such as DAMA/LIBRA and XENON100,
that seem  contradictory when interpreted in terms of WIMPs.

\section{Acknowledgments}

I thank  J.R. Cudell, for useful advice. This work is supported by
the Belgian fund F.R.S-FNRS.

\bibliographystyle{bibTex_styles/hope}
\bibliography{bibTex_styles/references}

\providecommand{\href}[2]{#2}\begingroup\raggedright\begin{thebibliography}{10}

\bibitem{Bernabei:2010mq}
{\bfseries DAMA/LIBRA} Collaboration, R.~Bernabei {\em et~al.}, ``{New results
  from DAMA/LIBRA},''
  \href{http://dx.doi.org/10.1140/epjc/s10052-010-1303-9}{{\em Eur. Phys. J.}
  {\bfseries C67} (2010) 39--49},
\href{http://arxiv.org/abs/1002.1028}{{\ttfamily arXiv:1002.1028
  [astro-ph.GA]}}.

\bibitem{Bernabei:2013xsa}
R.~Bernabei, P.~Belli, F.~Cappella, V.~Caracciolo, S.~Castellano, {\em et~al.},
  ``{Final model independent result of DAMA/LIBRA-phase1},'' {\em Eur. Phys.
  J.} {\bfseries C73} (2013) 2648,
\href{http://arxiv.org/abs/1308.5109}{{\ttfamily arXiv:1308.5109
  [astro-ph.GA]}}.

\bibitem{Aalseth:2011wp}
C.~Aalseth, P.~Barbeau, J.~Colaresi, J.~Collar, J.~Diaz~Leon, {\em et~al.},
  ``{Search for an Annual Modulation in a P-type Point Contact Germanium Dark
  Matter Detector},''
  \href{http://dx.doi.org/10.1103/PhysRevLett.107.141301}{{\em Phys. Rev.
  Lett.} {\bfseries 107} (2011) 141301},
\href{http://arxiv.org/abs/1106.0650}{{\ttfamily arXiv:1106.0650
  [astro-ph.CO]}}.

\bibitem{Collar:2013}
{\em Search for an annual modulation in 3.4 years of CoGeNT data}.
\newblock Talk presented at TAUP 2013, Asilomar, California, September, 2013.

\bibitem{Angloher:2011uu}
G.~Angloher, M.~Bauer, I.~Bavykina, A.~Bento, C.~Bucci, {\em et~al.},
  ``{Results from 730 kg days of the CRESST-II Dark Matter Search},''
  \href{http://dx.doi.org/10.1140/epjc/s10052-012-1971-8}{{\em Eur. Phys. J.}
  {\bfseries C72} (2012) 1971},
\href{http://arxiv.org/abs/1109.0702}{{\ttfamily arXiv:1109.0702
  [astro-ph.CO]}}.

\bibitem{Agnese:2013rvf}
{\bfseries CDMS-II} Collaboration, R.~Agnese {\em et~al.}, ``{Dark Matter
  Search Results Using the Silicon Detectors of CDMS II},'' {\em Phys. Rev.
  Lett.} (2013) ,
\href{http://arxiv.org/abs/1304.4279}{{\ttfamily arXiv:1304.4279 [hep-ex]}}.

\bibitem{Aprile:2012nq}
{\bfseries XENON100} Collaboration, E.~Aprile {\em et~al.}, ``{Dark Matter
  Results from 225 Live Days of XENON100 Data},''
  \href{http://dx.doi.org/10.1103/PhysRevLett.109.181301}{{\em Phys. Rev.
  Lett.} {\bfseries 109} (2012) 181301},
\href{http://arxiv.org/abs/1207.5988}{{\ttfamily arXiv:1207.5988
  [astro-ph.CO]}}.

\bibitem{Ahmed:2010wy}
{\bfseries CDMS-II} Collaboration, Z.~Ahmed {\em et~al.}, ``{Results from a
  Low-Energy Analysis of the CDMS II Germanium Data},''
  \href{http://dx.doi.org/10.1103/PhysRevLett.106.131302}{{\em Phys. Rev.
  Lett.} {\bfseries 106} (2011) 131302},
\href{http://arxiv.org/abs/1011.2482}{{\ttfamily arXiv:1011.2482
  [astro-ph.CO]}}.

\bibitem{Akerib:2013tjd}
{\bfseries LUX Collaboration} Collaboration, D.~Akerib {\em et~al.}, ``{First
  results from the LUX dark matter experiment at the Sanford Underground
  Research Facility},''
\href{http://arxiv.org/abs/1310.8214}{{\ttfamily arXiv:1310.8214
  [astro-ph.CO]}}.

\bibitem{Foot:2012rk}
R.~Foot, ``{Mirror dark matter interpretations of the DAMA, CoGeNT and
  CRESST-II data},'' \href{http://dx.doi.org/10.1103/PhysRevD.86.023524}{{\em
  Phys. Rev.} {\bfseries D86} (2012) 023524},
\href{http://arxiv.org/abs/1203.2387}{{\ttfamily arXiv:1203.2387 [hep-ph]}}.

\bibitem{Cline:2012is}
J.~M. Cline, Z.~Liu, and W.~Xue, ``{Millicharged Atomic Dark Matter},''
  \href{http://dx.doi.org/10.1103/PhysRevD.85.101302}{{\em Phys. Rev.}
  {\bfseries D85} (2012) 101302},
\href{http://arxiv.org/abs/1201.4858}{{\ttfamily arXiv:1201.4858 [hep-ph]}}.

\bibitem{Feldman:2007wj}
D.~Feldman, Z.~Liu, and P.~Nath, ``{The Stueckelberg Z-prime Extension with
  Kinetic Mixing and Milli-Charged Dark Matter From the Hidden Sector},''
  \href{http://dx.doi.org/10.1103/PhysRevD.75.115001}{{\em Phys.Rev.}
  {\bfseries D75} (2007) 115001},
\href{http://arxiv.org/abs/hep-ph/0702123}{{\ttfamily arXiv:hep-ph/0702123
  [HEP-PH]}}.

\bibitem{Khlopov:2010ik}
M.~Y. Khlopov, A.~G. Mayorov, and E.~Y. Soldatov, ``{The dark atoms of dark
  matter},'' {\em Prespacetime. J.} {\bfseries 1} (2010) 1403--1417,
\href{http://arxiv.org/abs/1012.0934}{{\ttfamily arXiv:1012.0934
  [astro-ph.CO]}}.

\bibitem{Khlopov:2011me}
M.~Y. Khlopov, A.~G. Mayorov, and E.~Y. Soldatov, ``{Towards Nuclear Physics of
  OHe Dark Matter},'' in {\em {Proceedings to the $14^{th}$ Workshop "What
  Comes Beyond the Standard Model"}}, N.~Mankoc~Borstnik, H.~B. Nielsen, C.~D.
  Froggatt, and D.~Lukman, eds., pp.~94--102.
\newblock 2011.

\bibitem{Cudell:2012fw}
J.~R. Cudell, M.~Khlopov, and Q.~Wallemacq, ``{The nuclear physics of OHe},''
  in {\em {Proceedings to the $15^{th}$ Workshop "What Comes Beyond the
  Standard Model"}}, N.~Mankoc~Borstnik, H.~B. Nielsen, and D.~Lukman, eds.,
  pp.~10--27.
\newblock 2012.

\bibitem{McCullough:2013jma}
M.~McCullough and L.~Randall, ``{Exothermic Double-Disk Dark Matter},''
\href{http://arxiv.org/abs/1307.4095}{{\ttfamily arXiv:1307.4095 [hep-ph]}}.

\bibitem{Fornengo:2011sz}
N.~Fornengo, P.~Panci, and M.~Regis, ``{Long-Range Forces in Direct Dark Matter
  Searches},'' \href{http://dx.doi.org/10.1103/PhysRevD.84.115002}{{\em
  Phys.Rev.} {\bfseries D84} (2011) 115002},
\href{http://arxiv.org/abs/1108.4661}{{\ttfamily arXiv:1108.4661 [hep-ph]}}.

\bibitem{2002ApJ...564...60M}
J.~Miralda-Escud{\'e}, ``{A Test of the Collisional Dark Matter Hypothesis from
  Cluster Lensing},'' \href{http://dx.doi.org/10.1086/324138}{{\em Astrophys.
  J.} {\bfseries 564} (2002) 60--64},
\href{http://arxiv.org/abs/0002050}{{\ttfamily arXiv:0002050 [astro-ph]}}.

\bibitem{Fan:2013yva}
J.~Fan, A.~Katz, L.~Randall, and M.~Reece, ``{Double-Disk Dark Matter},''
\href{http://arxiv.org/abs/1303.1521}{{\ttfamily arXiv:1303.1521
  [astro-ph.CO]}}.

\bibitem{Wallemacq:2013hsa}
Q.~Wallemacq, ``{Milli-interacting Dark Matter},''
  \href{http://dx.doi.org/10.1103/PhysRevD.88.063516}{{\em Phys. Rev.}
  {\bfseries D88} (2013) 063516},
\href{http://arxiv.org/abs/1307.7623}{{\ttfamily arXiv:1307.7623
  [astro-ph.CO]}}.

\bibitem{Amsler:2008zzb}
{\bfseries Particle Data Group} Collaboration, C.~Amsler {\em et~al.},
  ``{Review of Particle Physics},''
\href{http://dx.doi.org/10.1016/j.physletb.2008.07.018}{{\em Phys. Lett.}
  {\bfseries B667} (2008) 1--1340}.

\bibitem{Erkol:2005jz}
G.~Erkol, R.~Timmermans, and T.~Rijken, ``{The Nucleon-sigma coupling constant
  in QCD Sum Rules},'' \href{http://dx.doi.org/10.1103/PhysRevC.72.035209}{{\em
  Phys. Rev.} {\bfseries C72} (2005) 035209},
\href{http://arxiv.org/abs/0603056}{{\ttfamily arXiv:0603056 [nucl-th]}}.

\bibitem{McDermott:2010pa}
S.~D. McDermott, H.-B. Yu, and K.~M. Zurek, ``{Turning off the Lights: How Dark
  is Dark Matter?},'' \href{http://dx.doi.org/10.1103/PhysRevD.83.063509}{{\em
  Phys. Rev.} {\bfseries D83} (2011) 063509},
\href{http://arxiv.org/abs/1011.2907}{{\ttfamily arXiv:1011.2907 [hep-ph]}}.

\bibitem{Segre:1977}
E.~Segre, {\em {Nuclei and Particles}}.
\newblock W. A. Benjamin, Inc., New York, 2nd~ed., 1977.

\bibitem{Jaeckel:2010ni}
J.~Jaeckel and A.~Ringwald, ``{The Low-Energy Frontier of Particle Physics},''
  \href{http://dx.doi.org/10.1146/annurev.nucl.012809.104433}{{\em
  Ann.Rev.Nucl.Part.Sci.} {\bfseries 60} (2010) 405--437},
\href{http://arxiv.org/abs/1002.0329}{{\ttfamily arXiv:1002.0329 [hep-ph]}}.

\end{thebibliography}\endgroup

\end{document}